\documentclass[sigconf]{acmart}
\AtBeginDocument{%
  \providecommand\BibTeX{{%
    \normalfont B\kern-0.5em{\scshape i\kern-0.25em b}\kern-0.8em\TeX}}}

\copyrightyear{2024}
\acmYear{2024}
\setcopyright{acmlicensed}\acmConference[CHI '24]{Proceedings of the CHI Conference on Human Factors in Computing Systems}{May 11--16, 2024}{Honolulu, HI, USA}
\acmBooktitle{Proceedings of the CHI Conference on Human Factors in Computing Systems (CHI '24), May 11--16, 2024, Honolulu, HI, USA}
\acmDOI{10.1145/3613904.3642158}
\acmISBN{979-8-4007-0330-0/24/05}

\newcommand{\oursystem}{Fast-Forward Reality}

\usepackage{amsmath}
\usepackage{enumitem}

\newif \ifdraft \drafttrue   

\providecolor{added}{RGB}{31, 81, 255}   
\providecolor{deleted}{RGB}{255, 173, 176}     

\begin{document}

\title[\oursystem{}]{\oursystem{}: Authoring Error-Free Context-Aware Policies with Real-Time Unit Tests in Extended Reality}

\author{Xun Qian}
\authornote{Work completed when the first author interned at Meta Reality Labs.}
\email{qxziuan@gmail.com}
\orcid{0000-0003-1976-7992}
\affiliation{%
  \institution{Meta Reality Labs Research \& \\ Purdue University}
  \streetaddress{}
  \city{}
  \state{}
  \country{}
  \postcode{}
}
\author{Tianyi Wang}
\email{tianyiwang@meta.com}
\orcid{0000-0001-9382-6466}
\affiliation{%
  \institution{Meta Reality Labs Research}
  \streetaddress{}
  \city{}
  \state{}
  \country{}
  \postcode{}
}
\author{Xuhai Xu}
\email{xoxu@mit.edu}
\orcid{0000-0001-5930-3899}
\affiliation{%
  \institution{Meta Reality Labs Research \& \\ MIT}
  \streetaddress{}
  \city{}
  \state{}
  \country{}
  \postcode{}
}
\author{Tanya R. Jonker}
\email{tanya.jonker@meta.com}
\orcid{0000-0001-8646-5076}
\affiliation{%
  \institution{Meta Reality Labs Research}
  \streetaddress{}
  \city{}
  \state{}
  \country{}
  \postcode{}
}
\author{Kashyap Todi}
\email{kashyap.todi@gmail.com}
\orcid{0000-0002-6174-2089}
\affiliation{%
  \institution{Meta Reality Labs Research}
  \streetaddress{}
  \city{}
  \state{}
  \country{}
  \postcode{}
}
\renewcommand{\shortauthors}{Qian, et al.}

\begin{abstract}

Advances in ubiquitous computing have enabled end-user authoring of \emph{context-aware policies} (CAPs) that control smart devices based on specific contexts of the user and environment.
However, authoring CAPs accurately and avoiding run-time errors is challenging for end-users as it is difficult to foresee CAP behaviors under complex real-world conditions.
We propose \oursystem{}, an Extended Reality (XR) based authoring workflow that enables end-users to iteratively author and refine CAPs by validating their behaviors via simulated \emph{unit test cases}.
We develop a computational approach to automatically generate test cases based on the authored CAP and the user's context history.
Our system delivers each test case with immersive visualizations in XR, facilitating users to verify the CAP behavior and identify necessary refinements.
We evaluated \oursystem{} in a user study ($N$=12).
Our authoring and validation process improved the accuracy of CAPs and the users provided positive feedback on the system usability.

\end{abstract}



\begin{teaserfigure}
  \includegraphics[width=\textwidth]{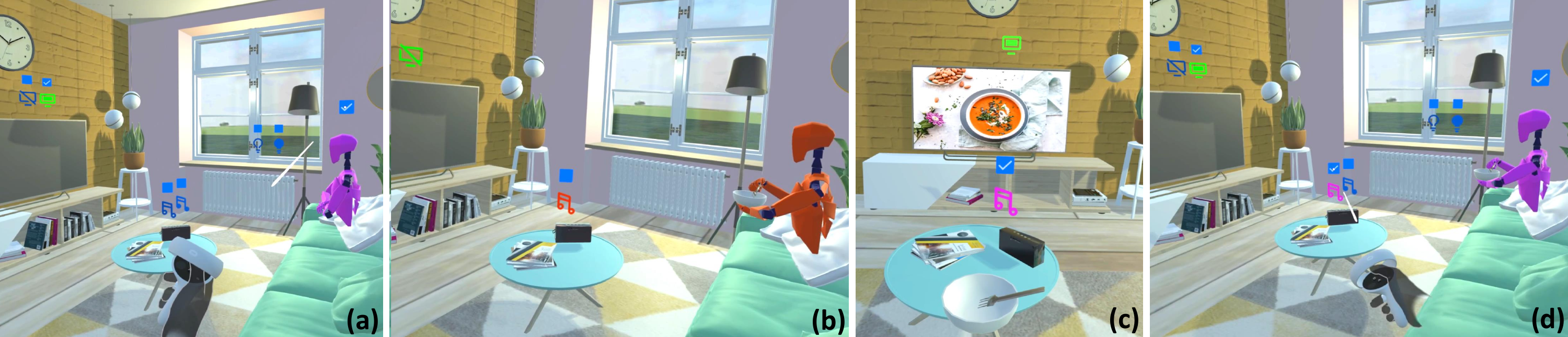}
  \caption{Overview of the \oursystem{} workflow: (a) An end-user initiates an authoring session of a context-aware policy (CAP) while being immersed in an Extended Reality (XR) authoring environment. The user includes \textit{context instance}s into the CAP by selecting the corresponding XR icons. (b) The system adaptively generates multiple \textit{unit test case}s based on the \textit{context history} of the user's everyday routines and the CAP, and highlights them with XR visualizations. (c) The user acts out the suggested \textit{test case}s by in-situ actions and selections of the involved \textit{context instance}s to intuitively validate whether the CAP reacts as expected under each \textit{test case}. (d) The user refines the CAP after noticing potential improvements and repeats the validation until the CAP is error-free under all \textit{test cases}.}
  \label{fig:teaser}
\end{teaserfigure}

\maketitle

\section{Introduction}

Developments in ubiquitous computing \cite{weiser1999computer} and smart environments \cite{coen1998design} have enabled automating functionality of Internet-of-Things (IoT) and smart devices. \emph{Context-aware policies} (CAPs) can define behaviors of these devices under personalized contexts of end-users and environments \cite{dey2001conceptual} (e.g., turn on smart lights after sunset).
With trigger-action programming \cite{ur2016trigger}, end-users can author customized CAPs by specifying context factors as triggers (e.g., `after sunset' and `leaving home'), and smart functions as actions (e.g., `turn on the lights' and `turn off the A/C').
To address diversified user needs, researchers and commercial products have greatly expanded the range of context factors to include a variety of environmental identities (e.g., time, temperature, weather, smart objects \cite{alexa,ifttt,dey2006icap, minaam2018smart,ye2023proobjar}), as well as user's location \cite{li2004topiary,wang2012location,ye2022progesar}, status \cite{zhang2005enabling}, and activities \cite{wang2020capturar,wu2013spatio}.
While most existing tools focus on enabling end-users to author CAPs with great flexibility, a \emph{runtime} issue emerges: it is challenging for users without programming skills to verify that the customized CAPs will behave as intended under varying contexts.

The following two scenarios illustrate how the authored CAPs may deviate from the user's original expectations.
The first CAP, \emph{``Dining'' $\to$ ``Turn on the TV''}, was authored on a relaxed evening, intended to automate an entertaining event while dining. However, this CAP is unexpectedly triggered another day when the user eats sandwiches while busily working in the study room.
The second CAP, \emph{``Doing yoga in front of the living room TV at night'' $\to$ ``Play the yoga playlist''}, was created by adopting the exact contexts during the authoring process. On another morning, the user expects music while doing yoga in the bedroom, yet, the authored CAP will not activate.
These examples highlight two types of inaccuracies that frequently occur in CAP authoring.
The first user forgot to consider the location and activity factors. So, the CAP is \emph{\textbf{under-specified}} and may trigger actions in unwanted scenarios (false positives).
Conversely, the second CAP is \emph{\textbf{over-specified}} since the user did not realize that redundant context factors were added so that the CAP could not be triggered in other desired scenarios (false negatives).
Such failures, if happen frequently, can lead to annoyance, frustration, loss of trust, and ultimately abandonment of such systems \cite{muir94trust}.
To avoid these errors, systems should provide users with intelligibility \cite{bellotti2001intelligibility} such that users can accurately predict the functionality of an authored CAP under various scenarios.
However, prior authoring tools and workflows have not investigated how they can address this crucial challenge. 
Hence, our work is motivated by the need for CAP authoring systems that assist end-users in \emph{proactively} eliminating potential inaccuracies in CAPs before deploying them.

In the software engineering industry, \textbf{\emph{unit testing}} \cite{zhu1997software} demonstrates a systematic approach to effectively avoid run-time errors \cite{griebe2014model,wang2014improving,sama2008model}. It is worth exploring whether this principled method could be used to ensure correctness in the context of end-user authoring of CAPs.
A high-quality \emph{unit testing} relies on a set of carefully constructed \textbf{\emph{test cases}} that is not only comprehensive enough to cover all corner cases but also precise enough with no redundancy.
Nevertheless, unlike expert programmers and developers, end-users lack the expertise. If asking users to manually design \textit{test cases}, they may be biased to the present contexts or consider contexts that are irrelevant to the current CAP \cite{ramakers2015paperpulse,todi16sketchplore}. To this end, we aim to assist common users in generating \textit{test cases} that are tailored to their local contexts, life routines, and the authored CAP. This way, users can efficiently validate the CAP in scenarios that are highly possible to happen in the runtime.

In addition, expert software engineers have the expertise to determine the expected outputs of each \textit{test case} and identify errors in the original function. In CAP authoring, considering that CAPs are deployed in end-users everyday environments (e.g., home, office), \textit{test cases} include diversified combinations of not only general digital contexts (e.g., current time, having a meeting) but also space-sensitive contexts that vary across each deploying environment (e.g., smart object states or human actions at different locations). 
For instance, in an environment with multiple smart lamps around several couches in the living room, it introduces ambiguity to describe \textit{test cases} that involve a specific smart lamp using typical text-based unit test tools \cite{pytest}. Further, it requires additional mental efforts for end-users to distinguish the imaginary contexts of a \textit{test case} from the present contexts. Hence, we aim to design an intuitive way to deliver each \textit{test case} so that non-expert users can effortlessly interpret it as a real scenario that may happen in the local environment, facilitating users to specify the desired performance and identify errors in the authored CAP.

With these major considerations, we propose \emph{\textbf{\oursystem{}}}, a novel system that enables users to validate and make corrections to CAPs during authoring by following the ``\emph{test-fault-correct}'' routine of unit testing.
\oursystem{} first introduces a computational framework that supports generating high-quality \textit{test cases} tailored to each user and the deploying environment. The framework organizes a user's context history perceived by intelligent technologies (e.g., computer vision, IoTs, head-mounted devices) as a series of contextual combination instances. By analyzing the frequency and correlation of different contexts, it understands the user's habits and preferences. Given an authored CAP, the framework can compose \textit{test cases} with specific combinations of related contexts where the CAP has a high possibility to happen in their lives but may perform inaccurately. In this way, the user can efficiently validate the CAP with highly customized \textit{test cases}.
Moreover, inspired by the prior works that immerse users in Extended Reality (XR) to facilitate the interpretation of space-sensitive contexts \cite{wang2020capturar,chae2016smart,grubert2016towards}, \oursystem{} adopts an XR interface that delivers each \textit{test case} by overlaying (e.g., smart object states), augmenting (e.g., time), and placing (e.g., human action) corresponding context visualizations in-situ in either the physical environment or the virtual replica. 
The user can intuitively evaluate whether the CAP performs correctly as if experiencing it in real life. Meanwhile, the XR interface serves as the main authoring interface to initiate an authoring session (\ref{fig:teaser}a) and iterate the CAP (\ref{fig:teaser}d). By repeating the \emph{author-test-refine} process, the user gradually removes the ambiguity of the CAP and becomes confident that the CAP will perform accurately once deployed.
In summary, the main contributions of this work are:

\begin{itemize}[leftmargin=*]
    \item An \emph{author-test-refine workflow} that enables end-users to validate and iterate CAPs at author-time by evaluating their performances via diverse simulated unit tests.
    
    \item A \emph{computational approach for generating unit test cases} that are personalized to each user and environment to effectively reveal potential run-time inaccuracies of the CAP.
    
    \item An \emph{XR-based authoring interface} that uses immersive visualizations to offer intuitive understandings of contexts presented in test cases, and direct operations to define and iterate the CAP.
\end{itemize}

\section{Related Work}\label{sec:relatedworks}

The main contribution of our work is a novel approach for end-users to author and validate CAPs in XR. To this end, we review the current literature on context-aware computing, end-user authoring, validation techniques, and immersiveness in XR.

\subsection{Context-Aware Policies for End-Users}

With the development of Internet-of-Things (IoT) devices \cite{li2015internet} and context-aware systems \cite{schilit1994context,abowd1999towards}, researchers have explored applications that automatically execute smart functions in everyday life when pre-specified contexts are detected \cite{dey2001understanding}.
We describe such automated applications as \emph{context-aware policies (CAPs)}. 

Following Dey's taxonomy of contexts \cite{dey2001conceptual}, four major types of human-centered contexts have been applied for such applications: \emph{identity}, \emph{location}, \emph{status(activity)}, and \emph{time}.
While \emph{time} and \emph{identity} (e.g., weather and calendar events) are straightforward contexts that can be easily sensed by computing systems \cite{dey2006icap, ifttt,alexa}, sensing location, activity, and user state is a more challenging problem.
Prior works have used external motion sensors (e.g., RFID \cite{weinstein2005rfid} and UWB \cite{oppermann2004uwb}), information-theory-based frameworks \cite{roy2003location}, or predefined privacy-preserving queries \cite{celdran2014secoman} to determine user location.
Beyond typical IoT applications, user location has been used to provide Ambient Assisted Living (AAL) services for the elderly \cite{helal2003enabling,mainetti2016iot}.
Activity detection has been explored using both external \cite{chiang2010interaction} and wearable \cite{wang2011recognizing} sensors to enable smart policies such as reminders and notifications based on user interactions (e.g., remind a resident that a specific food is about to expire during meal preparation).
Further, description logic rules \cite{wongpatikaseree2012activity}, and machine learning and deep learning approaches \cite{du2019novel,bianchi2019iot,wu2013spatio} have been applied to infer human activities from raw sensor data towards deploying activity-aware policies such as starting workouts or launching applications. 
Light and noise sensors have been used to infer other human states such as tiredness and whether a person is sleeping to automate smart home accessories such as lights and window blinds  \cite{zhang2005enabling}.
IMU sensors have been used for fall detection \cite{greene2016iot,kostopoulos2016f2d} to provide automatic AAL services.
Researchers have also proposed several approaches to perceive the status of IoT objects and leverage them into CAPs. 
Smart medicine boxes and bottles \cite{bhati2017smart, minaam2018smart}, water bottles \cite{borofsky2018accuracy}, and refrigerators \cite{floarea2016smart} are capable of counting and identifying their contents towards providing relevant CAPs such as smart reminders.

While many context-aware systems can provide generalized automatic services, the habits and needs of end-users vary in daily life.
Consequently, context factors and desirable policies are also highly dependent on personal preferences \cite{dey2001conceptual}. 
Thus, enabling end-users to \emph{author} CAPs that include user-defined contexts and personalized smart functions has been an area of great interest.

\subsection{End-User Authoring of Context-Aware Policies}

To help end-users author context-sensitive applications, `trigger-action' programming has been broadly adopted following the findings from an elicitation study \cite{dey2006icap}. 
Here, an end-user specifies a policy with a series of contexts as the trigger and a smart function as the action.
At run-time, when all the specified contexts are present, the system executes the corresponding action. 

Block-based programming in the style of `if this then that' \cite{ifttt,ur2014practical} has been widely used in commercial products such as Alexa routine \cite{alexa} and Apple Shortcuts \cite{shortcuts}. 
Here, a user selects contexts represented using 2D icons into the `this' block and adds smart functions into the `that' block.
Alternatively, the user can define such CAPs through sketches and descriptions \cite{dey2006icap}, programming-by-demonstration \cite{dey2004cappella,lee2013tangible}, or by segmenting maps for location-aware applications \cite{li2004topiary}.
Augmented Reality (AR) and Virtual Reality (VR) authoring techniques have recently been explored to provide users with spatial awareness and immersiveness.
CAPturAR \cite{wang2020capturar} supports users to author activity-based context-aware applications when referring to their past activities that are visualized using AR avatars. By defining proxemics with physical surroundings using a mobile-device-based AR authoring interface, ProGesAR \cite{ye2022progesar} and ProObjAR \cite{ye2023proobjar} enable users to include locations, object movements, and gestures into CAPs.
Further, Ivy \cite{ens2017ivy} immerses users into the digital twin of an indoor environment to define logic connections among smart objects via visual programming, while Haidon et al. \cite{haidon2020joining} enables a caregiver to author CAPs tailored to a resident's habits by selecting the remote resident's spatial-sensitive contexts that are mapped to the caregiver's local AR space using a semantic mapping approach.

Existing efforts have mainly targeted end-user challenges in authoring CAPs with specific context triggers and actions.
Some prior research has supported users to manually test authored CAPs by either selecting the triggering contexts on a 2D GUI \cite{dey2006icap, ramakers2015paperpulse} or by demonstrating the human activities and proxemics in AR \cite{wang2020capturar,ye2022progesar, ye2023proobjar}.
However, prior works do not investigate whether the user-authored CAPs perform according to user expectations in diversified run-time scenarios.
When such automation policies behave unexpectedly, users can lose trust and abandon these systems \cite{muir94trust}.
In this paper, we aim to address this issue by facilitating the validation and refinement of CAPs during authoring.

\subsection{Validation of Context-Aware Policies}

During real-world usage, end-users may encounter complicated contexts with diverse and unexpected edge cases \cite{de2017test,luo2020survey,matalonga2015challenges}. 
Thus, validating CAPs under such varied scenarios and resolving discrepancies has been widely explored in professional context-aware application development.
By adopting the approach of unit testing \cite{zhu1997software,hamill2004unit}, researchers have proposed multiple architectures and frameworks to help application developers design effective unit test cases to validate context-aware applications.

First, prior works propose data-driven approaches to generate diversified contextual scenarios to reflect the complex run-time conditions \cite{wang2007automated}. CASS \cite{park2007cass} proposes an architecture to automatically generate virtual sensory data for context-aware system developers to rapidly test and identify conflicts in the context-aware systems. TestAware \cite{luo2017testaware} allows developers to design test cases for mobile context-aware applications by downloading, replaying, and emulating contextual data on either physical devices or emulators.
On the other hand, it is impractical to validate countless real-world scenarios. 
The number of test cases should be constrained while maintaining the effectiveness of the validation. Context diversity \cite{wang2014improving,wang2010correlating} has been used to address this concern, where the distance between test cases is used to ensure diversity.
Other model-based approaches \cite{griebe2014model,yu2016testing,mirza2018automated,sama2008model} have been developed to efficiently generate test cases that can fulfill the validation requirements.
Similarly, this concern has been addressed in the research area of adaptive systems. ScalAR \cite{qian2022scalar} adopts the Genetic Algorithm to enable adaptive AR application authoring within a small scale of indoor layouts. It considers how to generate the most representative and diversified scenes with spatial and identity variations to help AR designers efficiently validate the AR element behaviors and identify failure cases. Inspired by these prior arts, we distill 3 major considerations for validating CAPs: 
(1) test cases should cover diversified and nuanced real-world scenarios, 
(2) test cases should be diverse but concise so that all potential scenarios can be captured with a limited number of test cases to ensure an efficient validation process, 
and (3) test cases should be able to effectively detect faults in applications.

Creating test cases that meet these criteria is a challenging task.
Unlike professional developers, end-users may lack the expertise to carefully design suitable test cases with varying complexity that cover as many corner cases as possible. Further, unlike general context-aware applications and programs, user-authored CAPs are highly personal; as such, test cases should be personalized to each user while addressing diverse scenarios. We believe that an automatic approach to generating personalized and diverse test cases can address these challenges.

Furthermore, to fully support novice users in validating and refining their CAPs, how to present these test cases appropriately remains challenging as a majority portion of the contexts perceived by advanced AI modules are associated with the spatial information (e.g., locations \cite{ye2022progesar}, proximity \cite{ye2023proobjar}, interactions \cite{wang2020capturar, huang2021adaptutar}, and IoT states \cite{huo2018scenariot}). However, the above-mentioned CAP validation tools do not address the need to enable local users to interpret the test cases precisely. Since the test cases are designed by professional developers who have a clear expectation of the outcome and solution to iterate the context-aware functions. Yet, when end-users author a CAP in their local environment without any expertise in designing the test cases, the test cases should be delivered to the users in an easy-to-understand manner. Two difficulties then pop up if using the existing tools: First, as we mentioned in Section 1, spatial-sensitive contexts often cause ambiguity when the referred assets are duplicated in space, especially when a test case may involve multiple contextual elements simultaneously. Prior research \cite{ballard1997deictic,bangerter2004using} has mentioned the inaccuracies caused by pure descriptive narrations. Meanwhile, since end-users do not have the expertise to design the test cases, feelings of non-confident and untrustworthy will be raised if no \emph{feedforwards} can be provided to end-users \cite{vermeulen2013crossing,vermeulen2012understanding,djajadiningrat2002Feedforward}.

Hence, in this paper, we endeavor to adopt an approach that can intuitively represent the test cases to end-users, and more importantly, facilitate the users to rapidly identify the errors in the existing CAPs by thoroughly interpreting the provided test cases.

\subsection{Immersiveness in XR}

Motivated by prior research, we believe Extended Reality (XR) could provide an effective medium to communicate test cases for validation to users.
Enabled by spatial awareness of XR, prior works have used virtual augmentations adjacent to relevant physical entities to facilitate end-users to understand complicated contexts such as human activities \cite{wang2020capturar,cao2019ghostar}, interactions \cite{cao2020exploratory,huang2021adaptutar,wang2021gesturar,liu2023instrumentar}, and status of smart objects \cite{huo2018scenariot}.
Further, while being immersed in XR environments, end-users can experience simulated scenarios free from the limitations of physical contexts such as user status and time.
Users can revisit what happened in the past via in-situ XR visualizations of human activities \cite{fender2022causality,lindlbauer2018remixed,cao2019ghostar} and interactions \cite{qian2022arnnotate,xia2018spacetime}, or virtually immerse into other times and spaces \cite{qian2022scalar,wang2021distanciar,zhu2023learniotvr}.
Finally, through immersive authoring \cite{lee2004immersive}, users can intuitively program and author XR applications from within the target deployment environment \cite{nebeling2021xrstudio, nebeling2020xrdirector,zhang2020flowmatic,he2023ubi}, which enables a real-time and seamless transition between authoring and testing.

To this end, our work develops an XR authoring environment that enables users to (1) iteratively author a CAP, (2) immersively evaluate auto-generated unit test cases with complex contextual scenarios, and (3) directly manipulate test cases and observe simulated results to obtain feedback on whether a CAP performs as expected and make refinements as needed.
\section{\oursystem{}}

We present \oursystem{}, a novel XR-based workflow that enables end-users to iteratively author accurate CAPs through unit testing. 
To this end, we develop a computational approach for generating diverse and personalized unit test cases and an authoring interface that provides users with effective CAP validation via immersive visualizations.
In this section, we define a set of design goals, that guide the system development, and present a walkthrough to illustrate the ``author-test-refine'' workflow for a specific CAP.
Following this, we elaborate upon our technical approach and system implementation.

\subsection{Design Goals} \label{sec:design-goals}

In this work, we aim to develop a system that helps end-users author accurate CAPs via real-time validation before deployment. 
While being immersed in the XR authoring environment, a user can experience multiple unit test cases that are personalized to their previous activities and routines.
Through simulated \emph{feedforwards}, the system provides intelligibility and enables refinement when outcomes do not match original expectations.
Consequently, users can eliminate potential runtime inaccuracies that may surface during initial authoring.
To effectively support users in the authoring and refinement process, we postulate that a system should fulfill the following \textbf{design goals}.

\begin{figure*}[h]
  \centering
  \includegraphics[width=\linewidth]{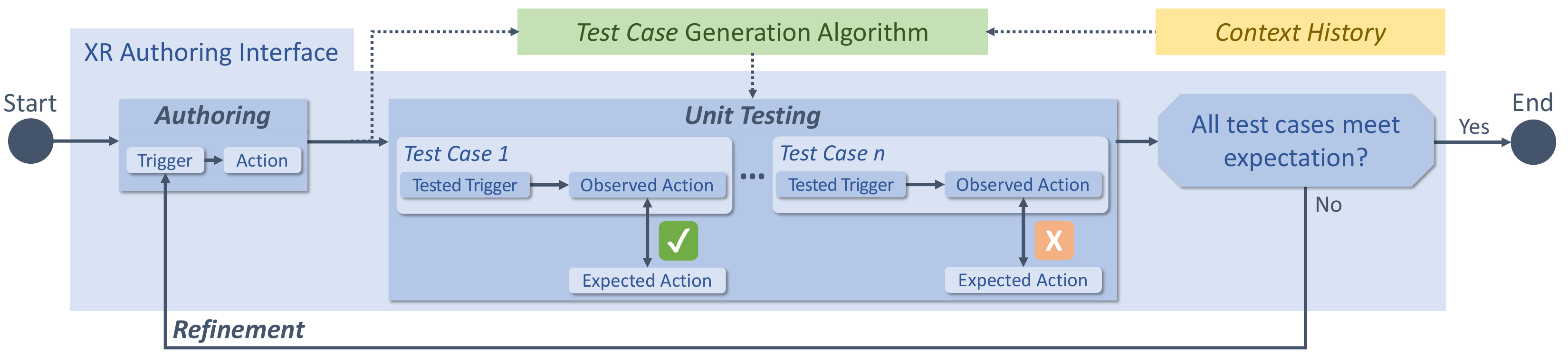}
  \caption{The authoring workflow of \oursystem{}. While being immersed in the authoring environment, a user starts \emph{authoring} to define a target action and initial \textit{context instance}s. 
  The system generates test cases from the user-authored CAP and context history. When the user starts \emph{unit testing}, test cases are visualized to enable validation of the CAP's behavior.
  If unexpected instances are identified, the user \emph{refines} the original CAP by editing the \textit{context instance}s. The user repeats this process to iteratively refine the CAP until it meets their expectations.}
  \Description{}
  \label{fig:workflow}
\end{figure*}

\begin{enumerate}[leftmargin=*]
    \item \textbf{\textit{Personalization}}: As addressed in prior CAP authoring systems \cite{dey2006icap,dey2004cappella,wang2020capturar}, test cases should be closely related to the CAP a user is authoring and personalized to their context history. This enables the user to effectively refine CAPs by examining contextual scenarios that are likely to occur in their everyday life.
    
    \item \textbf{\textit{Diversification}}: Prior works \cite{luo2017testaware,park2007cass} have addressed the need to diversify the test cases by proposing different frameworks and guidelines. Following their suggestions, to cover a wide range of possible scenarios, test cases should be diverse and capture edge cases that may cause unexpected outcomes. 
    
    \item \textbf{\textit{Brevity}}: Test cases should be concise such that users can clearly identify factors that cause inaccuracies. Further, the number of test cases should be constrained such that they are tractable within a limited time \cite{wang2014improving,wang2010correlating}.

    \item \textbf{\textit{Interpretability}}: As we discussed above, since end-users may not have professional programming expertise, it is cumbersome to interpret a combination of multiple space-sensitive contexts involved in a test case. And hence, it is difficult for end-users to imagine possible iterations of the CAPs. Thus, we propose to leverage the advantages of XR so that users can intuitively understand the complicated contexts and immersively experience the test cases that may happen in the future.
    
    \item \textbf{\textit{Seamlessness}}: While being immersed in the XR environment, the interactions required for authoring and validating a CAP should be fluent and consistent to ensure seamless transitions in the \textit{author-test-refine} workflow following the findings of prior immersive authoring and spatial programming works \cite{lee2004immersive,nebeling2021xrstudio,zhang2020flowmatic,ens2017ivy}.
\end{enumerate}

\subsection{Target Scenarios and System Walkthrough}

In this paper, we identify the research scope where an increasing number and types of contexts can be detected and included in context-aware systems through modern AI modules (e.g., object detection, activity recognition), IoT communication, and AR spatial awareness. Prior works \cite{dey2006icap,dey2004cappella,li2004topiary,wang2020capturar,ye2022progesar} have explored various systems to integrate different contexts into context-aware systems. We acknowledge their contributions and assume that it is straightforward to integrate these modules into a CAP authoring system. We follow example scenarios that are similar to the one shown in Figure \ref{fig:teaser} to illustrate the system design of \oursystem{} in the next sections. Specifically, a user could live in an apartment with a layout similar to Figure \ref{fig:teaser}. The user regularly wears a head-mounted XR device with real-time object detection and activity recognition capabilities. Meanwhile, smart objects such as smart lights, TVs, music players, and coffee makers are present in this space, while calendar events, dates, and weather information are intrinsically available in their XR devices. 

Figure \ref{fig:workflow} illustrates the workflow of \oursystem{} in a target scenario as mentioned above. Here, we explain it using the example shown in Figure \ref{fig:teaser}, where an end-user tries to author a CAP to control a smart TV.
The user lives in a studio apartment and prefers to watch TV while eating.
One day, as the user is watching TV while sitting on the sofa, the user decides to author a CAP to automatically turn on the TV during such situations.

\textbf{Authoring}: The user activates \oursystem{} and begins the authoring process.
By selecting in-situ icons that represent `TV is on' and `location is sofa' \textit{context instance}s, the user authors an original CAP: \emph{`turn on the TV when I am on the sofa'} (Figure \ref{fig:teaser}a).

\textbf{Unit Testing}: Then, they start the validation process to ensure correctness.
A computational algorithm examines the user's context history and the current CAP, and identifies that the user always `eat' (activity) while `watching TV', but rarely engages in other activities such as `reading' or `using the phone'. However, in the current CAP, the user has not included any activity \textit{context factor}.
Additionally, the context history indicates that when the `TV is on', the `music player' is typically `off'.
However, this case is not considered in the CAP either.
Therefore, the system generates a \emph{test case}: `eating on the sofa while the music player is off', and visualizes it in the XR environment (Figure \ref{fig:teaser}b).
The user interacts with the immersive visualization and enacts the actions of sitting on the sofa and eating, while the music is off, and observes the TV turning on (Figure \ref{fig:teaser}c).

\textbf{Refinement}: Following the highlighted relevant contexts in the \emph{test case}, the user identifies their importance and \emph{refines} the under-specified CAP by adding the two suggested \textit{context instance}s (Figure \ref{fig:teaser}d).
The CAP is now more accurate such that the TV is not turned on when the user is engaged in other activities or listening to music. And, the user keeps validating more system-generated \emph{test cases} based on the updated CAP until no refinements are needed.
By using \oursystem{}, the user has successfully authored and verified an accurate CAP to turn on their smart TV under desirable contextual scenarios.

\subsection{Framework for Authoring} \label{sec:context-history-record}

\begin{figure*}[t]
  \centering
  \includegraphics[width=\linewidth]{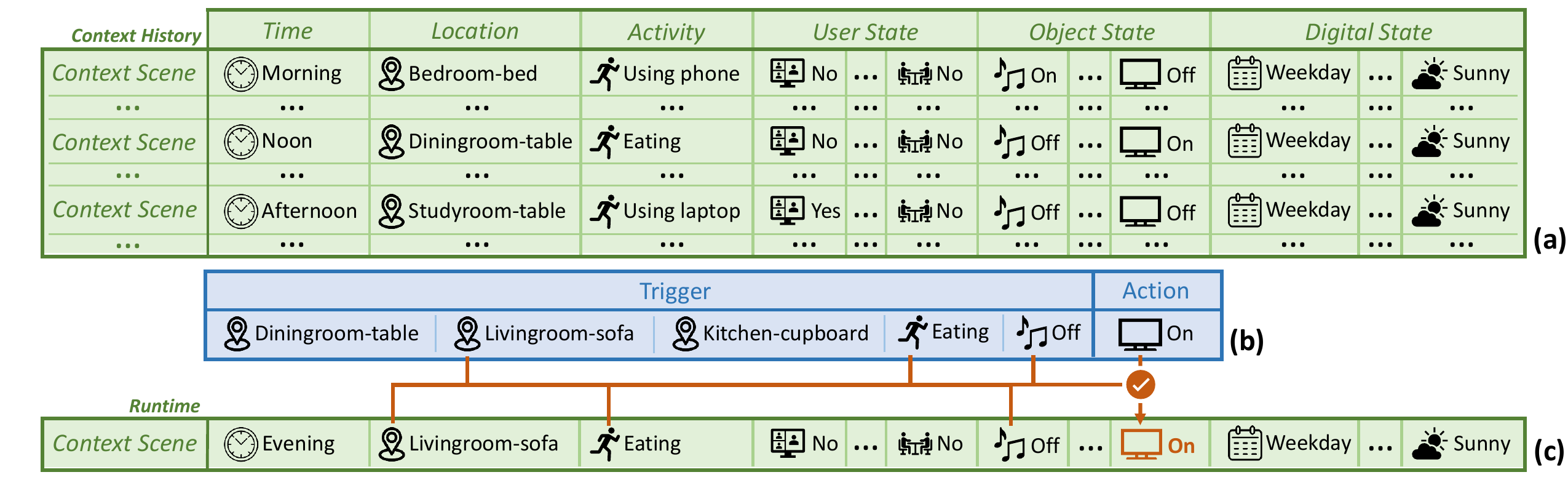}
  \caption{The framework of authoring CAP using \oursystem{}. (a) A user's everyday contexts are recorded as the \textit{context history}, which contains a series of \textit{context scene}s where each one is a collection of concurrently occurring \textit{context instance}s of corresponding \textit{context factor}s that can be detected in the environment. (b) In a user-authored CAP, the `trigger' contains multiple \textit{context instance}s from the same or different \textit{context factor}s, while the `action' is a \textit{context instance} of a smart object that reflects the functional state (e.g., `TV is on'). (c) During deployment, for all the \textit{context factor}s included in the CAP, when the specified \textit{context instance}s are present in the real-time \textit{context scene}, the `action' is triggered.}
  \Description{}
  \label{fig:framework}
\end{figure*}

In this section, we elaborate on the framework adopted by the initial authoring and its connections with the context perception techniques. It addresses three concerns: (1) following prior research, how does \oursystem{} detect and structure a user's everyday contexts, (2) during authoring, how does the user define the components of a CAP, and (3) upon deployment, how does our system determine when should the action of the CAP be executed.
As discussed in Section \ref{sec:relatedworks}, technical solutions have been developed to detect and identify increasing types of contexts that were proposed in early elicitation works \cite{schilit1994context,abowd1999towards,dey2006icap}. 
We adopt previously proposed taxonomy of contexts \cite{abowd1999towards,dey2001conceptual} and assume integration with modern AI modules to categorize the contexts we aim to leverage into the following \textbf{\textit{context factor}}s:

\begin{itemize}[leftmargin=*]
    \item \textbf{\emph{Time}} represents the time of a day, which is commonly used in CAPs: `turn on all lights in the `evening'. It is the most straightforward context that can be retrieved.
    
    \item \textbf{\emph{Location}} of a user serves as a critical factor in a CAP (e.g., `turn on the A/C when I enter the living room').
    We assume that the user locations can be detected by external sensors \cite{oppermann2004uwb,weinstein2005rfid} or through computer vision with advanced head-mounted devices \cite{hololens2} in real-time.
    
    \item \textbf{\emph{Activity}} belongs to the \textit{status(activity)} category \cite{dey2001conceptual}, representing both general activities and interactions with objects that are detectable via software \cite{wongpatikaseree2012activity} and hardware \cite{zhang2019sozu} based activity detection modules. Considering the structure of currently available activity detection networks and benchmarking datasets \cite{cao2017realtime,damen2018scaling,grauman2022ego4d}, we assume activities will be detected as discrete labels (e.g., eating, reading, etc.) given continuous sliding window time series of human skeleton or hand-object interaction as inputs.
    
    \item \textbf{\emph{User state}} also belongs to the \textit{status(activity)} category \cite{dey2001conceptual}. It represents the spatial-insensitive status of a user. Typical examples include: `being alone', `in a meeting', and `feeling tired'. These contexts can be detected via AI networks or inferred from other contexts (e.g., how many cups of coffee have been drunk). Typically, the outputs are discrete nominal labels.
    
    \item \textbf{\emph{Object state}} represents status of objects in the environment. For a smart object such as a coffee machine or a smart pill bottle, its physical states (e.g., not enough water, how many pills left) can be identified but not digitally manipulated. For other smart and IoT devices, their active status can be altered by a CAP (e.g., `turn on the TV'). Using IoT communication protocols such as Resource Description Framework \cite{decker2000semantic}, and hardware-based localization algorithms such as ultra-wideband \cite{huo2018scenariot}, such discrete contexts can be detected during runtime.
    
    \item \textbf{\emph{Digital state}} is related to \textit{identity} and represents general but unique environmental information provided in general digital devices such as \textit{temperature} and \textit{weather}. Similar to the \textit{Time}, these contexts can also be easily retrieved from the HMD.
\end{itemize}

For each type of \textit{context factor}, we define \textbf{\textit{context instance}}s as the nominal values (labels) that can be assigned to them. 
For instance, the location factor can either include object-oriented \textit{context instance}s such as `living room sofa' and `dining table', or non-interpretable anchor IDs that are marked by AR devices. 
Activity can include labels from activity detection models (e.g., `sleeping', `walking').
Boolean `yes' and `no' states are used to indicate object states (`music player is on').
At any moment in a user's daily life, we define a \textbf{\textit{context scene}} as a combination of \textit{context instance}s, where each \textit{context instance} belongs to one available \textit{context factor} (Figure \ref{fig:framework}a). 
Whenever one of the \textit{context factor}s changes, a new \textit{context scene} is registered, and all the recorded \textit{context scene}s form the user's personal  \textbf{\textit{context history}} (Figure \ref{fig:framework}a).

To enable end-user CAP authoring, we adopt the widely-used trigger-action programming paradigm \cite{ur2016trigger}. 
Here, a user first defines an `action' to manipulate a target object (e.g., \textit{TV is on}), and then defines a `trigger' by including multiple \textit{context factor}s (e.g., \textit{time} and \textit{location}), and for each \textit{context factor}, selecting the \textit{context instance}s (e.g., \textit{morning} and \textit{afternoon}) (Figure \ref{fig:framework}b).
During runtime, when a CAP is deployed, for all included \textit{context factor}s, if one \textit{context instance} is present in the current \textit{context scene}, the corresponding smart function (`action') is executed (Figure \ref{fig:framework}c).

\subsection{\textit{Test Case} Generation in Unit Testing}

\begin{figure*}[h]
  \centering
  \includegraphics[width=\linewidth]{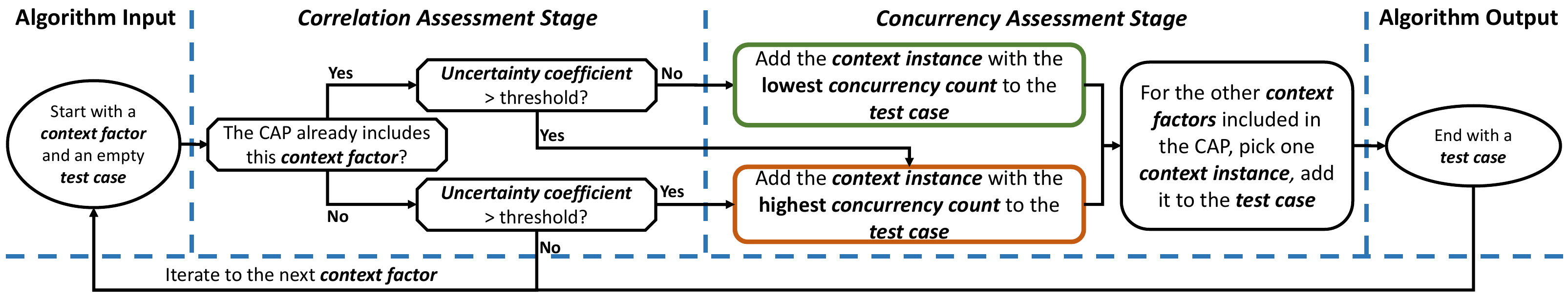}
  \caption{\textit{Test case} generation algorithm. For each \textit{context factor}, the algorithm starts with an empty \textit{test case}. In the \textit{correlation assessment stage}, we investigate whether the processing \textit{context factor} holds a high \textit{uncertainty coefficient} value with the target action, and whether it is already selected into the CAP. Depending on the conditions, the algorithm then processes the \textit{concurrency assessment stage} to select the \textit{context instance}s into the \textit{test case}. In one specific condition, the system will skip this process and directly iterate to the next \textit{context factor}, while in other conditions, the algorithm outputs a \textit{test case} and starts to process the next \textit{context factor}.}
  \Description{}
  \label{fig:algo}
\end{figure*}

After a user has created an initial CAP, the main purpose of the unit test stage is to help reveal potential errors (the under-specified and over-specified scenarios discussed before), which are caused by the inaccurate selection of the contexts of interest following the framework introduced before. Note that resolving runtime errors of AI modules is out of this work's scope. Typically, most of the prior CAP authoring \cite{dey2006icap,wang2020capturar,dey2004cappella} and validation \cite{luo2017testaware,wang2014improving,yu2016testing} systems assume the detection of the contexts is reliable and robust. For instance, an alarm CAP can be validated by changing a simulated time. However, it is impossible to test and refine the CAP when the time itself is wrong. Thus, the \textit{test case} generation algorithm we propose in this paper only aims to resolve the inaccuracies in the user-authored CAPs, rather than the unpredictable AI errors. Yet, we will discuss the potential extension in the Discussion section.

With the above CAP authoring framework, we follow the unit testing approach and develop a computational approach to generate diverse \textbf{\textit{test case}}s that help reveal potential inaccuracies in a CAP. A \textit{test case} consists of a set of \textit{context instance}s available in the environment.
Our \textit{test case} generation algorithm addresses the design goals of \emph{personalization}, \emph{diversification}, and \emph{brevity} (defined in Section \ref{sec:design-goals}) through three key strategies:

\begin{itemize}[leftmargin=*]
    \item \textbf{\textit{Strategy A - Correlation}}: Investigate which \textit{context factor}s largely affect the execution of the current CAP and include appropriate \textit{context instance}s into \textit{test case}s. By doing so, when visiting the \textit{test case}, the user can notice the scenarios that are highly associated with their personal preferences.
    
    \item \textbf{\textit{Strategy B - Intention}}: Add one (and only one) \textit{context instance} for each \textit{context factor} that is already in the CAP so that the \textit{test case} is closely related to the user's original idea and realistically reflect a real-world condition where only one \textit{context instance} of a \textit{context factor} can happen at one moment.
    
    \item \textbf{\textit{Strategy C - Simplification}}: Exclude irrelevant \textit{context factor}s to avoid distracting the user so that they can focus on the \textit{context factor}s likely to cause errors.
\end{itemize}

To implement \textit{Strategy A}, as we assume that users will maintain some routines during everyday life, we compute \textbf{\textit{uncertainty coefficient}} \cite{theil1970estimation} to identify \textit{context factor}s.
Specifically, this is a conditional information entropy-based approach to asymmetrically reveal that in the presence of a particular \textit{context instance}, what is the probability that another \textit{context instance} occurs. 
For instance, a user may watch TV at different times of the day, but is always eating and not using the phone while doing so.
In this case, the \textit{uncertainty coefficient} between the `TV state' and `time' is low, while `activity' is high.
To apply this in our system, once the user selects a target action, we loop over every detectable \textit{context factor}, and (1) retrieve all past \textit{context instance}s of both the target action and the current processing \textit{context factor} from the user's context history and (2) calculate the \textit{uncertainty coefficient} between the two lists using the $theils\_u$ function\footnote{https://github.com/shakedzy/dython/blob/master/dython/nominal.py}.
Further, since these coefficients only reflect the \textit{context factor} level correlations, we also calculate the \textbf{\textit{concurrency count}} for each \textit{context instance} within the \textit{context factor}: (1) fetch all \textit{context scene}s where the target action \textit{context instance} happened and (2) construct a dictionary of all the \textit{context factor}s that saves the counts of \textit{context instance}s that happen in these \textit{context scene}s using the $Counter$ function\footnote{https://docs.python.org/3/library/collections.html\#collections.Counter}.
These \textit{concurrency count}s will help the system determine which \textit{context instance}s to include in the test case after deciding the \textit{context factor}.

Figure \ref{fig:algo} illustrates the \textit{test case} generation algorithm. Given the target action and the initial \textit{context instance}s, together with the pre-calculated \textit{uncertainty coefficient}s and \textit{concurrency count}s, we loop over every \textit{context factor} that is available in the current environment to process the algorithm.
First, in the \textbf{\textit{correlation assessment stage}}, we examine (1) whether the \textit{context factor} is already included in the CAP and (2) whether the \textit{uncertainty coefficient} of the currently processing \textit{context factor} is greater than an empirically set threshold (we will explain it in Implementation section).
Four conditions will be available and we only process 3 of them in the following \textbf{\textit{concurrency assessment stage}}.

\begin{enumerate}[leftmargin=*]
    \item If a \textit{context factor} shows a high correlation with the target action, however, is not included in the current CAP, we initiate a new test case and add the \textit{context instance} that most frequently happens concurrently with the action in the test case. Typically, this case helps eliminate under-specified mistakes. For instance, a user always does specific activities while `TV is on', leading to a high \textit{uncertainty coefficient}. However, the user has not included any activity in the CAP. So, the system includes the activity that the user mostly does when watching TV in the test case to remind the user of this critical factor in the CAP.
    
    \item If a \textit{context factor} shows a low correlation with the target action, however, is already included in the current CAP, we initiate a new test case and add the \textit{context instance} that rarely happens together with the action into the test case. Consider the same user, who would watch TV at different times, adds `evening' into the CAP. The system warns the user about this potentially over-specified error by including `time' when the user rarely watches TV in the test case.
    
    \item If a \textit{context factor} shows a high correlation with the target action, and is included in the current CAP, we initiate a new test case if there is a \textit{context instance} that happens more frequently. This condition also targets over-specified CAPs. For example, consider location holds a high \textit{uncertainty coefficient} and the user has already included the dining table in the CAP. However, \textit{concurrency count}s indicate that the user watches TV often while on the `sofa'. Our system includes `sofa' into the test case to nudge the user to generalize this \textit{context factor} in the CAP.
    
    \item Note that the last condition (i.e., a \textit{context factor} with low \textit{uncertainty coefficient} that is not included in the CAP) will not lead to any test cases because we assume that this decision made by the user already meets their personal preferences.
\end{enumerate}

Next, to realize \textit{Strategy B}, for each CAP-involved \textit{context factor} that is not included in the \textit{test case}, the system randomly picks one \textit{context instance}, such that the generated \textit{test case} is consistent with what the user just authored.
After iterating over all the $N$ \textit{context factor}s, the algorithm generates $M<N$ \textit{test case}s, where $M$ depends on the previously mentioned threshold. Meanwhile, if the current CAP involves $n$ \textit{context factor}s, each \textit{test case} involves $m \in \{n,n+1\}$ \textit{context instance}s. 
Finally, \textit{Strategy C} is satisfied by composing \textit{test case}s of only the most critical \textit{context instance}s that may affect the user's decision and omitting low-priority instances.
The user can focus on testing out the suggested \textit{test case}s without needing to pay attention to irrelevant \textit{context factor}s.
By adopting this principled approach, the system can generate multiple \textit{test case}s and present it to the user.

\subsection{XR Authoring Interface} \label{sec:authoring-interface}

\begin{figure*}[h]
  \centering
  \includegraphics[width=\linewidth]{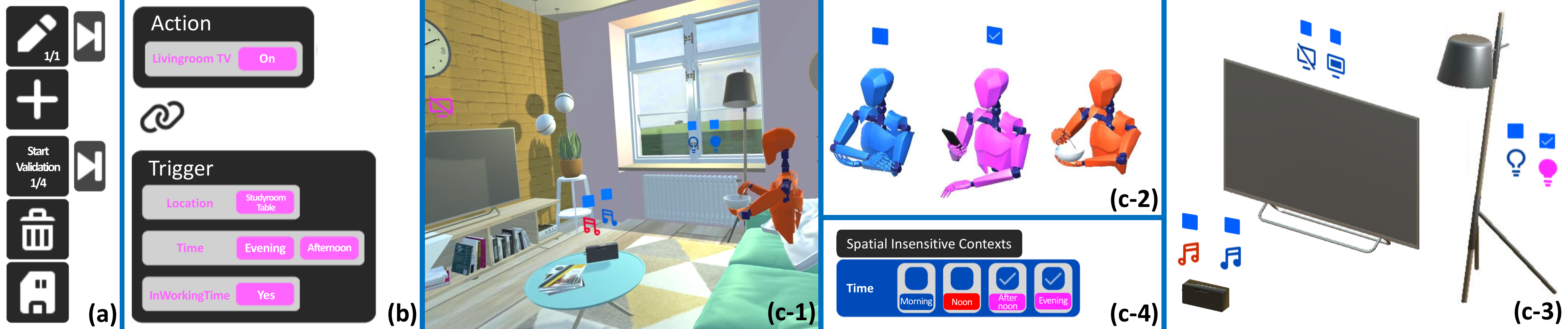}
  \caption{The XR-based authoring interface of \oursystem{}. (a) The main menu rendered on a user's non-dominant hand enables users to `edit' existing CAPs, 'add' a new CAP and author it, `start validation' of unit \textit{test case}s, `delete', and `save' the CAP. (b) An authoring panel displayed on the user's non-dominant hand indicates the action and triggers in the current CAP. (c-1) The immersive XR authoring environment. (c-2) Location and activity \textit{context instance}s are represented using avatars with different poses. (c-3) Spatial states are placed in situ, next to corresponding smart objects. (c-4) The digital state and user state are rendered on the user's hand.
  All \textit{context instance}s are color-coded to represent different conditions where blue color indicates that the \textit{context instance} is not selected, pink represents that it is included in the current CAP, and red represents that it is included in the current \textit{test case}.}
  \label{fig:interface}
  \Description{}
\end{figure*}

Our XR authoring environment ensures that \oursystem{} follows the design goals (\autoref{sec:design-goals} of providing \textit{interpretability} during testing and enabling \textit{seamlessness} of authoring and refinement.
In our system, a menu is rendered on the user's non-dominant hand encapsulating all buttons for controlling the system (Figure \ref{fig:interface}a). The `edit' and `edit next' buttons allow for editing one of the existing CAPs.
The `new' button initiates the authoring of a new CAP. The `start validation' button is used for switching between authoring and validation. Finally, the `save' and `delete' buttons allow users to store or remove a CAP.

When the user starts authoring a CAP, available \textit{context instance}s are displayed in the XR authoring environment (Figure \ref{fig:interface}c-1).
Users can interact with these elements to include them in the authored CAP.
Based on the types of \textit{context factor}s, we place spatial-sensitive \textit{context instance}s directly in the XR environment.
Specifically, the location and activity are illustrated using avatars with different poses (Figure \ref{fig:interface}c-2); object states are represented using XR icons placed next to the corresponding objects (Figure \ref{fig:interface}c-3); spatially-insensitive contexts (time, digital state, and user state) are displayed on the user's non-dominant hand with 2D icons (Figure \ref{fig:interface}c-4).
The user can select the checkbox above each \textit{context instance} to add/remove it to/from the CAP.
Meanwhile, each \textit{context instance} is color-coded\footnote{Additional textual labels can be included to ensure visual accessibility.} to indicate different selection states: we use blue to represent un-selected \textit{context instance}s, pink for those included in the CAP, and red for those in a \textit{test case}.
Lastly, a panel is displayed on the user's non-dominant hand (Figure \ref{fig:interface}b) to illustrate the \textit{context instance}s and activities included in the current CAP.

\textit{Test case}s are generated automatically and made available to the user.
When the user starts validation, a \textit{test case} is visualized within the same environment to enable quick interpretation.
Users can observe a \textit{test case} as a `fast-forward' simulation that reveals the consequences of a particular context scenario, thus providing them with \emph{feedforward}.
Typically, as illustrated in Section 3.4 and Figure 3, \oursystem{} compares the real-time detection of the AI modules with the user-authored CAP, and visualizes the corresponding output in the XR environment. For instance, if all the conditions are met for a TV controller, the physical TV will be turned on in the environment.
Since authoring and testing take place within the same environment, the user can immediately start refining their CAP by modifying the \textit{context instance}s as needed.
Note that to reduce users' visual and mental loads, all the XR UIs are only accessible to users during the author-test-refine flow. When users return to everyday life, \oursystem{} will hide all XR visualizations but keep running at the backend.

\subsection{Implementation} \label{sec:implementation}

We use Meta Quest 2 \cite{quest2} as the target platform for developing the authoring interface, and implemented the system using Unity3D (2020.3.16f1) \cite{unity}. Interactions with the interface are currently supported via handheld controllers; however, it is trivial to switch to free-hand interactions supported by Quest 2 and other XR devices \cite{hololens2}.
To determine a proper threshold value used in the \textit{test case} generation algorithm, and a suitable number of \textit{context scene}s to capture in the context history, we conducted preliminary tests using a context history collection tool, which is described in the next section. 
Typically, we generated $5, 10, 15, 20$ \textit{context factor}s that may be relevant to a home environment. 
For each case, we collected $10, 20,...,50$ \textit{context scene}s. 
During data collection, we followed an abstract rule to control smart functions and collected diverse \textit{context scene}s with a small amount of noise. 
Then, we calculated the \textit{uncertainty coefficient}s of the smart function concerning all available \textit{context factor}s.
We empirically set the correlation value to be $0.5$, then set the minimum number of \textit{context scene}s collected in the context history to be $10$ times the number of available \textit{context factor}s in the environment. 
By doing so, we ensure that the number of \textit{context factor}s that are higher than the threshold will be approximately less than $5$. 

\section{User Study}

To evaluate whether the design of \oursystem{} fulfills our design goals (Section \ref{sec:design-goals}), we conducted a systematic user study.
In this work, we propose an XR-based workflow that addresses the difficulties of validating complicated CAPs that cannot be straightforwardly solved by traditional developer-oriented unit testing systems. In specific, since end-users do not have the expertise to design high-quality test cases while in the local environments, we design a computational algorithm that leverages the users' personal history to extract diverse test cases. Moreover, we adopt the immersiveness enabled by the XR technique to show the test cases via in-situ virtual content and support users to intuitively understand each test case by enacting each test case in XR via in-situ simulation. Therefore, the goals of this study include: (1) investigating whether the \textit{test case}s generated by the system effectively meet users' needs in identifying potential inaccuracies of the CAPs, and (2) evaluating whether the XR-based environment helps users easily understand the test cases and identify potential improvements of the CAPs. Note that the main research scope of this work is not to compare whether our system outperforms prior approaches, instead, we aim to illustrate that the system can succeed in solving the research difficulties when dramatically increased spatial-sensitive contexts can be involved in CAP authoring.

\subsection{Study Setup}

For the study, we designed a virtual one-bedroom apartment (Figure \ref{fig:userstudy-setup}a) as the home environment to conduct the user study.
We also developed an interactive tool (Figure \ref{fig:userstudy-setup}b) to enable end-users to elicit personal \textit{context history} that is appropriate to the virtual home.

\subsubsection{Context History Collection Tool}

Our authoring process relies on the availability of a user's \textit{context history}, containing \textit{context scene}s that have taken place in the past.
Given the constraints of our controlled study and privacy concerns, collecting participants' real \textit{context history} was infeasible.
To circumvent this issue, we developed a data collection tool where participants could quickly explicate their everyday activities to create their \textit{context history} in the provided virtual home.
We asked participants to specify \textit{context instance}s temporally by prompting them to imagine their daily routines. 
Note that one of the key observations that has sparked HCI interest and research on developing CAPs is that humans do follow some routines and have personalized preferences in their everyday lives. Thus, we aimed to provide an environment that resembled each user's personal living environment and asked them not to deliberately create random non-representative context histories, but to follow some general personal routines.

\begin{figure}[h]
  \centering
  \includegraphics[width=\linewidth]{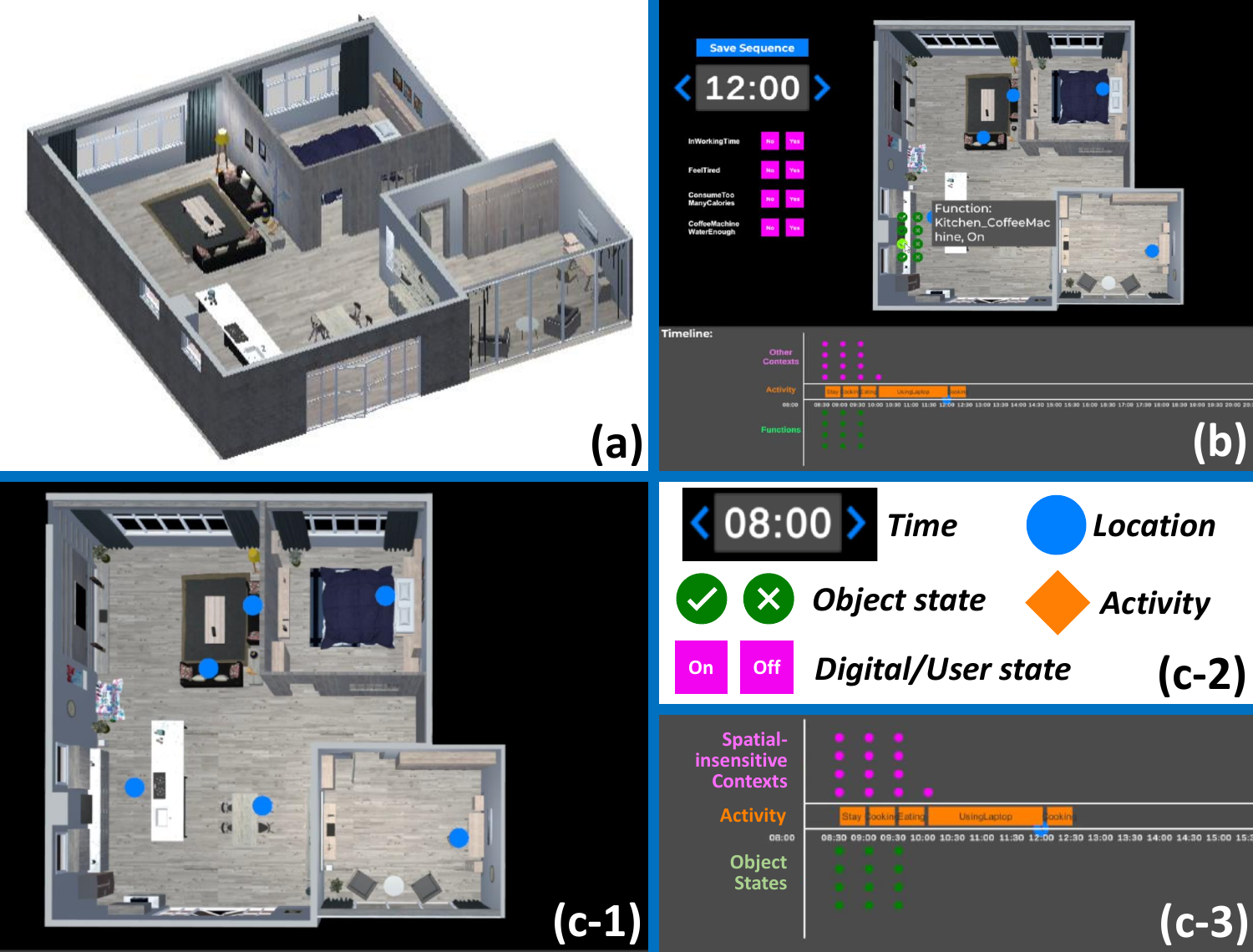}
  \caption{(a) The virtual environment used for the user study. (b) The \textit{context history} collection tool. The top left corner below the `new/save sequence' button contains spatial-insensitive contexts, \textit{time} and user-specific \textit{digital/user state}s (e.g., `is working', `feel tired', `consume too many coffees', and `not enough water in the coffee machine') (c-1) The floor plan of the target environment. By clicking each blue dot, pre-designated spatial-sensitive \textit{context instance}s such as \textit{object state}s and \textit{activities} can be toggled together with their \textit{location}s (e.g., `coffee machine is on/off', `stay', `cooking', and `eating'.) (c-2) The available \textit{context instance}s that are visualized in different icons according to the \textit{context factor}s. Typically, location is spatially overlaid on the floor plan, while available activities and object states are displayed as lists after selecting the location. (c-3) The timeline indicating all the recorded \textit{context instance}s. Explanation of the \textit{context instance} will appear when hovering on the icon. One can delete a \textit{context instance} by clicking the block/dot.}
  \Description{}
  \label{fig:userstudy-setup}
\end{figure}

As shown in Figure \ref{fig:userstudy-setup}b, the data collection interface is separated into four parts.
In the top left corner of the UI, a `new sequence' button allows users to create a new `day' of living in the virtual apartment, while a clock illustrates the time. 
After pressing the `new sequence' button, a user can select the two arrows next to the clock to change the time (Figure \ref{fig:userstudy-setup}c-2); the numeric time is encoded into nominal values (e.g., evening).
In the center, the floor plan (top-view) of the virtual apartment is displayed (Figure \ref{fig:userstudy-setup}c-1). 
All available location \textit{context instance}s are represented with selectable blue circular nodes. 
By selecting a node, the available \textit{context instance}s of the two types of \textit{context factor}s, activity, and object state are shown as lists. 
The participant can click an activity or an object state to indicate what will happen at the specified time.
Further, available user states are listed below the clock for the user to select.
All selectable \textit{context instance} nodes are illustrated in Figure \ref{fig:userstudy-setup}c-2; a textual description of the \textit{context instance} appears when the cursor hovers over the corresponding node.
User-specified \textit{context instance}s are visualized along a timeline (Figure \ref{fig:userstudy-setup}c-3).
The participant can click a block/dot to delete the corresponding \textit{context instance}.
After specifying all \textit{context instance}s at a specified time, the participant can change the time and repeat all operations until the `current day' is finished. 

Once the user presses the `save sequence' button, for every time-step when the user specifies at least one \textit{context instance}, a new \textit{context scene} is created and stored. If the user does not specify any \textit{context instance} of particular \textit{context factor}s, they are set to default values in the \textit{context scene}.
By consecutively eliciting multiple days, the user successfully provides a \textit{context history}.

\subsection{Study Method}

One of the major research questions we aim to investigate is the efficacy of \oursystem{} and whether validation and refinement at author-time can improve the accuracy of CAPs before deployment. 
To inform our system design, we derived a set of five design goals.
We conducted a within-subject comparative study to assess whether the test case generation algorithm can generate effective tests that fulfill the requirements of the first design goals 1--3 (\textit{personalization}, \textit{diversification}, \textit{brevity}).
Design goals 4--5 (\textit{interpretability}, \textit{seamlessness}) were assessed via a subjective questionnaire and interview.

Since the main goal was to prove the effectiveness of our \textit{test case} generation algorithm, we designed a baseline system that adopted the same XR authoring interface, while turning off the \textit{test case} generation capability. In specific, we provided the users with the same XR UIs as illustrated in Section 3.5. The only controlled factor was that during the test and refine phases, the users could manually toggle different \textit{context instance}s and spatial movements to validate their CAPs. Note that, the users had full freedom to stop testing if they felt the authored CAP was already precise enough.
In the study, each participant was asked to author two CAPs using \oursystem{} and the baseline system respectively in a virtual apartment (Figure \ref{fig:userstudy-setup}a).

\textbf{Participants:} 
12 participants were invited to the user study (8 males, 3 females, and 1 non-binary; mean age=27.83, SD=3.16). 
No specific background or experience was required during the recruitment. 
9 participants had used smart objects (e.g., Nest Thermostats and Philips Hue), and 3 out of these 9 had authored everyday automation (e.g., Alexa) at least once.
11 participants had previously used XR applications (e.g., Pokemon Go and Beat Saber), while 5 users had developed XR applications. 
None of the participants had seen or used our system before the study.

\textbf{Procedure:}
In the author-test-refine workflow, users do not complete the authoring in one shot, but continue modifying their CAPs until they feel satisfied with all the system-provided test cases. Thus, the most effective way to evaluate our system is to investigate the performance of the CAPs validated using the test cases.
Overall, each user was asked to use \oursystem{} and the baseline system to author 2 CAPs: (1) a CAP that controls a smart TV in the living room of the virtual apartment and (2) a CAP that controls a smart music player that can play music in the entire apartment.
To provide the users with a more realistic feeling of living in the virtual apartment, we sent a pre-study survey to the users before they came, which included (1) the study background, (2) the virtual home environment floor plan (Figure \ref{fig:userstudy-setup}a), (3) the 2 target smart functions, and (4) the available types of \textit{context factor}s. 
Then, the survey asked the users what \textit{context factor}s and \textit{context instance}s would be considered when they wanted to control the two smart functions. 
We then prepared the virtual apartment for both the 2D tool and the XR authoring environment accordingly.
Each user signed a consent form stating that no reimbursement was provided and the user preserved the right to terminate the study whenever they wanted.
After a user arrived, the researcher introduced the background and definition of CAPs, and the user collected the \textit{context history} using the 2D interface.
Then, the user entered the XR authoring environment to complete the 2 authoring tasks using the 2 systems while considering data counterbalancing among all the users.
During each task, we recorded the final authored CAP, completion time, number of iterations of the CAP (if a user added more than one \textit{context instance} at one time, it was still marked as one iteration), and number of \textit{test case}s created/validated.
After each task, the user was asked to complete a Likert-type survey towards the feedback on the validation process.
At the end of the study, the user completed another Likert-type questionnaire regarding the XR authoring environment and the entire usage experience, together with a Standard Usability Scale survey. 
Meanwhile, a conversation-like interview was conducted to collect the user's subjective feedback on the system.

\subsection{Study Results}

Based on the pre-study survey results, we included 8 types of \textit{context factor}s in the virtual apartment. 
Time, location, activity, TV state, and Music player state were available for all the users, while `is working', `is alone', `feel tired', `weather', `having a meeting', and `playing video games' were provided for different users based on their survey responses. 
Meanwhile, the available activities also varied (e.g., `doing exercise' and `smoking' were provided for some users).
During the collection of the \textit{context history}, we collected M=81.9 (SD=10.96) \textit{context scene}s, and selected 75\% of the data for the \textit{test case} generation algorithm and 25\% for accuracy evaluation discussed later.

We first report the analysis of the users' operations during the authoring processes. Note that these statistics and observations do not explicitly address our design goals, but provide implications of the usefulness of our system.
Overall, the users finished each task using the baseline system with 5.0 (SD=1.86) minutes, while 6.25 (SD=1.82) minutes using \oursystem{}. Within the entire authoring time, 2.25 (SD=1.60) minutes were spent for manually validating the CAPs, and using our system, 3.58 (SD=0.90) minutes were taken.
On average, the users generated 1.25 (SD=1.06) \textit{test case}s. We observed that using the baseline system, most users simply checked the CAP by selecting the \textit{context instance}s they added during the validation stage. 
Using \oursystem{}, 5.25 (SD=1.66) \textit{test case}s were evaluated by the users. Note that although approximately 4 times more \textit{test case}s were viewed, validation time was not greatly impacted.
This can be attributed to the rationale that with the baseline system, users spent more time thinking about what \textit{context instance}s could be used to validate the CAPs.
Furthermore, the users conducted 1.0 (SD=1.28) iterations of the CAPs after the manual validation to add more constraints to the CAPs. We also observed that the users tried to make the CAPs more accurate by adding more \textit{context instance}s even before the validation.
\textit{``I used IFTTT before, this was how I created a rule by just adding those important factors. It's always better to add more constraints to make it more accurate, right?'' (P6)}
In contrast, with the help of our system, 3.92 (SD=1.44) iterations were performed. Typically, we observed not only the `addition' of the \textit{context instance}s, but also the `removal' of the \textit{context factor}s after visiting some \textit{test case}s.
While these generalized observations partially substantiate our proposed approach, in the following analysis, we further investigated the accuracy of the CAPs and the users' subjective feedback systematically.

\subsubsection{\textit{Test case} Generation Evaluation}

\begin{figure*}[h]
  \centering
  \includegraphics[width=\linewidth]{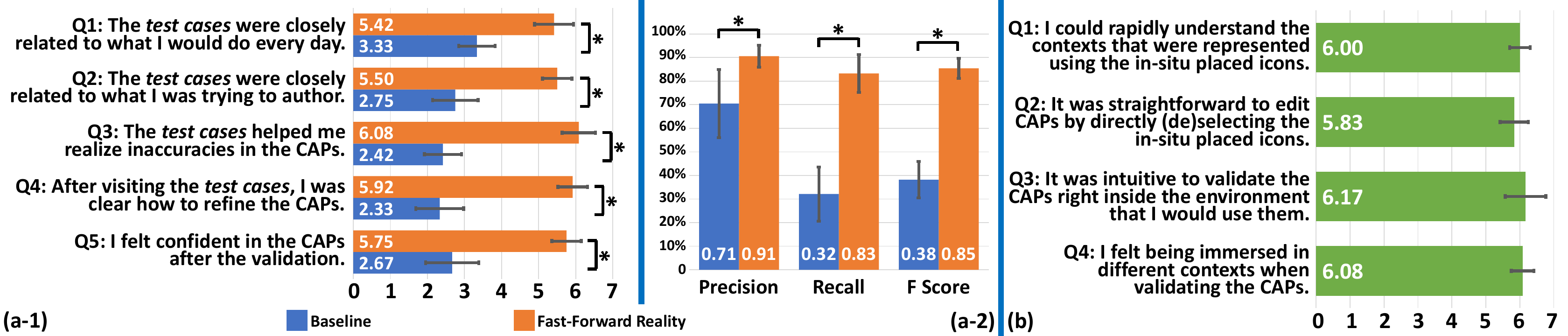}
  \caption{(a-1) The survey results of the quality of the \textit{test case}s generated by the users using the baseline system and by \oursystem{}. (a-2) The accuracy results of the CAPs authored by the users using the two systems. (b) The subjective feedback on the XR authoring environment.}
  \Description{}
  \label{fig:userstudy-results}
\end{figure*}

We first report the qualitative feedback on the \textit{test case} generation algorithm from the 7-point Likert-type questionnaire. 
Typically, the users were asked to answer the same set of questions after experiencing \oursystem{} and the baseline system respectively. The questions and results are shown in Figure \ref{fig:userstudy-results}a-1.
Besides the general analysis, we conducted a Wilcoxon Signed-Rank Test to investigate whether the feedback on the \textit{test case}s was significantly different between the two systems. This test approach specifically targeted within-subject non-parametric data while no normality test was needed.

As noted in the \textit{personalization} design goal, the \textit{test case}s should be closely related to the users' everyday activities (Q1) and the currently authoring CAPs (Q2). As shown in the results, our system was positively welcomed by the users in the relevance to the daily actions of the \textit{test case}s (M=5.4, SD=0.79), and showed a significant difference (Z=-2.8, p=0.004) compared with the baseline system (M=3.3, SD=1.44).
\textit{``One [\textit{test case}] where I ate food in the study room was definitely what I would do. It was one of my personal preferences.'' (P4)}
Regarding the relevance to the CAP, our system received a more decent rating (M=5.5, SD=0.80), compared with the baseline system (M=2.8, SD=1.29). A significant difference was identified in this question (Z=-3.1, p=0.002).
\textit{``Using the [baseline system], I had no idea of how to pick more [\textit{context instance}]s since the only idea I had was already added in the policy. But the [\oursystem{}] could give me some good examples, such as `having a meeting', and `working on my laptop in the music player task', which was definitely what I would care about.'' (P3)}
Another key motivation when generating \textit{test case}s is whether they could help users realize potential mistakes in the CAPs (the \textit{diversification} and \textit{brevity} design goals). 
The users highly appreciated that the \textit{test case}s could contribute to the refinement against the CAPs (Q3: M=6.1, SD=0.90, and Q4: M=5.9, SD=0.79). These ratings were significantly higher than the baseline system (Q3: M=2.4, SD=1.00 and Q4: M=2.3, SD=1.23): for Q3, Z=-3.1, p=0.002, and for Q4, Z=-3.1, p=0.002.
\textit{``I liked the idea of showing some other potential conditions of one trigger. For the music player, when I first authored it, I only added `evening' because I was thinking about eating dinner, but [\oursystem{}] also reminded me of cooking in the morning, which would also happen.'' (P11)}
\textit{``The system not only informed me which factor I should consider, but I could also directly use that case in my policy, such as it suggested me adding not working and not listening to music into the TV policy.'' (P1)}
\textit{``What made me surprised was that your system led me to think more about my policy. When I saw that eating at the dining table case, I started to think maybe I should add doing workouts in the living room to the TV policy as well.'' (P7)}
Last but not least, the users felt more confident after they validated the CAPs using \oursystem{} (Q5: M=5.8, SD=1.06), and the baseline system received a significantly lower rating (M=2.7, SD=0.98) with Z=-3.0, p=0.003.
\textit{``Actually, only after I used your system did I realize how bad it was when I created that music player rule using the [baseline system].'' (P2)}

Besides the users' subjective feelings, we investigated whether the performance of the CAPs was improved after the validation. Given a user-authored CAP, we used the rest of the \textit{context history} as the ground-truths to calculate: (1) precision = $TP/(TP+FP)$, (2) recall = $TP/(TP+FN)$, and (3) F-score = $2\cdot precision \cdot recall/(precision+recall)$, where $TP$ represents `true-positive' indicating the target action is expected to be executed while the CAP successfully triggers it; $TN$ represents `true-negative' where the CAP accurately stays silent when the target action should not be triggered; $FP$ (false-positive) means the CAP mistakenly trigger the target action but the user does not need it; $FN$ (false-negative) means the action should be executed, but the CAP does not work.
We also conducted a paired t-test after the data passed the normality test for the significance analysis (Precision: |Kurtosis|=0.88<2.0 and |Skewness|=0.36<2.0; Recall: |Kurtosis|=0.88<2.0 and |Skewness|=0.36<2.0; F-score: |Kurtosis|=0.88<2.0 and |Skewness|=0.36<2.0). 
The results are shown in Figure \ref{fig:userstudy-results}a-2. 

Using our system, the precision of the CAPs reached 90.6\% (SD=0.01), which was significantly higher than that of using the baseline system (70.5\%, SD=0.08): t(11)=-2.51, p<0.05.
Similarly, the recall was significantly increased from 32.1\% (SD=0.05) to 83.3\% (SD=0.03) with t(11)=-7.50, p<0.005.
Further, the overall F-score was improved from 38.2\% (SD=0.02) to 85.4\% (SD=0.01) with t(11)=-10.10, p<0.005.
The statistical analysis indicated that by using our system, the users could author more accurate CAPs via the additional validation stage.
Meanwhile, we noticed that using our system, the recall and F-score were greatly increased if compared with the precision. According to the definition of precision and recall, the numbers of the FN reduced to a greater extent than that of the FP when using \oursystem{} to validate the CAPs. One main reason was that when using the baseline system, no assistance was provided for the validation, thus many users added additional constraints that made the CAPs more over-specified.
On the other hand, when using our system, we observed that after visiting some \textit{test case}s, the users not only added more \textit{context instance}s to the CAPs, but also removed some \textit{context factor}s that were originally included.
\textit{``When using that [baseline system], there was only one scenario in my mind, so I just created that specific CAP When I used [\oursystem{}], those suggestions let me realize the CAP was too constrained.'' (P3)}
Another important observation was that although the overall accuracy was significantly increased, the CAPs rarely reached 100\% accuracy. It was because in everyday life, a person could not strictly follow one single CAP as the routine. There always exist scenarios that have never happened before. We will discuss it in the Limitation section in terms of how to address this issue.

\subsubsection{XR Authoring Environment Evaluation}

To address the design goals 4--5 (\textit{interpretability} and \textit{seamlessness}), \oursystem{} leverages XR to build an immersive authoring environment for end-users to experience \textit{test case}s and iterate the CAPs. 
Using a 7-point Likert-type survey, we evaluated whether the users welcomed the features of the authoring interface. The results are illustrated in Figure \ref{fig:userstudy-results}b.

We received complimentary feedback on the feature allowing users to try out the \textit{test case}s while being immersed in the XR environment (M=6.1, SD=0.67).
\textit{``Because I'll use these policies in the physical environment, it's a good idea to let me try out my [CAPs] in the same environment. When I looked at those virtual icons right above the TV and sofa, I could easily understand what they meant.'' (P4)}
Meanwhile, all the users agreed that it was necessary to validate the CAPs before the real-world deployment (M=6.2, SD=1.19).
\textit{``I felt much more confident about my CAP when I could see the TV was on after I sat on the sofa. That instant feedback was realistic.'' (P5)} Such comments were aligned with the findings in prior works in \emph{feedforward} simulation \cite{djajadiningrat2002Feedforward, vermeulen2013crossing}, where users tend to visualize the realistic outcomes instead of descriptions to gain trust against the digital applications.
Placing virtual icons that represent the corresponding affordance and functionalities of the smart objects received positive feedback (M=6.0, SD=0.60).
\textit{``For those spatial contexts, it's better to show them in the environment. Otherwise, if I have many smart lights, it's difficult to understand which light I'm referring to using pure texts.'' (P2)} Users also commented on the comparison between in-situ icons and 2D icons for spatial-insensitive contexts. \textit{``I wonder if those icons can also be attached to the corresponding objects. For example, the `having-meeting' context can be attached to my calendar. Then, I can fully focus on the physical world when I do the test.'' (P9)}
In addition, the immersive operations provided by our system were also highly accepted by the users (M=5.8, SD=0.83).
\textit{``Because the icons were inside the environment, I didn't have to switch between different platforms to create my CAPs.'' (P10)} Several users who had programming experience (P11, P7, and P1) also pointed out that our system was aligned with the trend of spatial programming \cite{zhang2020flowmatic,wang2020capturar} for enabling non-experts to join the application development. \textit{``If the program will be used in 3D, it is more intuitive to create and test it in 3D as well. It is a promising way to attract people who do not have coding expertise to create such CAPs in the future.'' (P11)}
Last but not least, a decent Standard Usability Scale score with M=86.0 and SD=6.77 further proved the overall usability of \oursystem{}.

\subsection{Discussion}
\oursystem{} has been proven effective in facilitating non-expert users to author error-free CAPs via properly generated \textit{test case}s in XR. Given the current scope of the CAP authoring, we discuss more insights and concerns we have distilled from the design process and user study results that inspire future research along the novel author-test-refine workflow of the CAP authoring area.

\subsubsection{Scalability of the system and validated CAPs}
In the current implementation, we illustrate our system in a virtual environment where spatial-sensitive \textit{context instance}s and corresponding UIs are placed in situ. On one hand, our system could easily adapt to the corresponding AR environment. Specifically, when introducing the framework of context awareness in Section 3.3, we clarified that the scope of this paper lies in the assumption that users live in AI-powered smart environments so that all types of the \textit{context factor}s could technically be registered and detected in the physical environment. While the registration is out of this paper's scope, prior works \cite{clark2022articulate,huo2018scenariot} have addressed this need. Hence, for all \textit{context instance}s and UI icons that require in-situ visualization (e.g., an activity happening at a sofa), \oursystem{} could display them at the corresponding physical locations. Meanwhile, \textit{object state}s and \textit{digital state}s are also available to display in the physical environment or anchored to a fixed place (e.g., coffee maker status, TV status, calendar events). 

Moreover, we would discuss the possibility of deploying one CAP across different environments, which has been addressed by prior authoring systems \cite{ifttt, qian2022scalar, yang2014verifying}. Following the current \textit{context factor} framework, one straightforward answer is that the CAP could be correctly run as long as all the involving \textit{context instance}s are present in the new environment.
Prior work \cite{qian2022scalar} enables designers to validate the AR application performance by providing scenes where corresponding spatial and semantic associations are absent. Such an idea could be leveraged in the \textit{test case} generation process by showcasing scenarios when each \textit{context instance} is not available.
However, the participants raised concerns when the researchers asked about more scenarios they would use the authored CAPs. \textit{``Actually, I was thinking of using the music CAP when I went back to my parent's home as well. But I realized it would be a lot different. I may create a different CAP.'' (P6)} \textit{``It depends on my preference. Some CAPs are just for home, I don't even want to turn it on somewhere else. But some CAPs, like more general ones, I would expect that to work all the time.'' (P12)}
As CAPs are highly associated with humans' daily routines, any addition and deletion of \textit{context instance}s represent the specific intention. Therefore, how to balance between scalability and personalization for authoring CAPs still requires further studies.

\subsubsection{Mixed errors beyond CAP authoring}
The main contributions of this work lie in the validation workflow and the algorithm to generate high-quality unit \textit{test case}s that help identify and eliminate over/under specified errors.
However, while researchers in the AI domain have been continuously working on improving the AI model performance and robustness, detection errors are inevitable under varied scenarios \cite{li2017artificial}, and this issue has also been explored by other context-aware system research \cite{lim2010toolkit,yang2014verifying}.
In this paper, the user study has proven that \oursystem{} could effectively avoid user-caused errors during CAP definition. It was conducted in a virtual environment and the accuracy of the CAPs was calculated by running the CAPs in the user-collected \textit{context scene}s where all the \textit{context instance}s were assumed to be correctly detected. Yet, we recognize the importance of bridging the gap between the validation process and the imperfect performance of the current AI technologies. Users would lose trust in our system if a CAP performs wrongly due to an object detection mistake even though the logic of the CAP is properly validated using our system.
Intrinsically, the AI errors do not belong to the logic of any CAP. Thus, it is impractical to directly introduce such uncertainty into the validation workflow, as users could refine nothing logically on the CAP even though they know it may fail due to AI mistakes.
However, we propose two potential methods to mitigate the mixed error issue for our system and future CAP authoring systems.
First, sensing errors could happen during the \textit{context history} collection, which reduces the correctness of the \textit{test case} generation algorithm. We could enable users to eliminate wrong detection data so that the generated \textit{test case}s are entirely reliable. Inspired by CAPturAR \cite{wang2020capturar}, we could allow users to revisit the past activities and contexts with XR visualizations, then either delete or correct the wrong \textit{context history}. 
Further, to deal with the sensing errors during execution, we notice that Explainable AI (XAI) has become a popular topic that allows for explaining AI errors to end-users to increase user trust in AI \cite{arrieta2020explainable}. 
We envision an integration between our system and an XAI agent that can pop up the list of detected contexts when a CAP does not perform as expected to inform the user that such a mistake either comes from the CAP authoring or AI detection mistakes \cite{lim2010toolkit,antifakos2005towards}.

Another uncertainty resides in the users' daily behaviors. A person never strictly follows a routine in complicated everyday life. Although our system improved the CAP accuracy via validation, the F-score of the CAPs could never reach 100\%.
One way to further improve the accuracy could be authoring multiple CAPs and adopting XAI approaches as discussed previously.
Additionally, we envision the user wearing an advanced XR-HMD all the time for the detection of the contexts. Thus, the specific \textit{context scene} that causes the mistake could be either recorded as a corner case and treated separately in the future or used for improving the AI-based models. 
We could also improve the \textit{test case} generation algorithm by taking into consideration these \textit{context scene}s that cause deployment errors.

\subsubsection{Effectiveness of XR interfaces in CAP authoring and validation}
As discussed in Related Works, 2D-based interfaces are prevalent in commercial CAP authoring tools and professional context-aware application development. With the advances of the XR technology, only a few works \cite{wang2020capturar} explored the advantages of utilizing this emerging technology in CAP authoring tasks.
In this paper, the positive feedback received on the innovative XR-based testing approach, along with XR's immersive capabilities, suggests that the symbiosis of XR and CAP remains a promising area for future research. 
On one hand, the spatial awareness of XR contributes to a more intuitive representation of spatially sensitive contexts such as \textit{object state}s and \textit{activities}, which empowers users to intuitively include contexts from the attaching physical objects or locations. Further, the immersive visualization of different \textit{context instance}s enables users who do not have expertise in the concept of unit testing to rapidly understand imaginary contextual scenarios, and identify failures. This \emph{feedforward} idea supported by XR successfully bridges the temporal gap between authoring and usage of a CAP, and enables the novel author-test-refine workflow introduced in this paper.
We believe that the fusion of XR technology with CAP authoring not only enriches the process with greater intuitiveness and immersion but also facilitates understanding complicated real-world contextual combinations during authoring. This synergy, therefore, merits further research exploration, promising significant advancements in the field of CAP development.
Additionally, unlike conventional 2D-based UIs which have undergone extensive refinement over the decades, the design of XR interfaces still needs large-scale and long-term studies to lower the user friction associated with this relatively new technology. In the next section, we outline various concerns that could guide further improvements of the XR interfaces in CAP authoring.

\section{Limitation and Future work}

\textbf{More complicated contexts and CAPs.}
In this work, we primarily focus on the authoring of one single CAP following the framework discussed in Section \ref{sec:context-history-record}. 
The quantitative results of the user study proved that with the additional validation stage, the users could successfully iterate the CAPs to make them more accurate.

One concern is that the current design of \oursystem{} follows the mainstream CAP authoring tools that leverage nominal context factors (e.g., discrete labels output by Machine Learning modules). Yet, prior works such as CAPturAR \cite{wang2020capturar} and DART \cite{macintyre2004dart} have adopted the programming by demonstration metaphor to enable the definition and detection of non-nominal contexts (e.g., a `clip' of human activity, a time series of audio signals that represent people having group conversations). We envision integrating similar features in \oursystem{} where users could directly demonstrate specific \textit{context factor}s rather than selecting them one by one. However, how to generate test cases that help identify errors requires further research. For instance, how the system knows the semantics of a demonstrated `clip' and finds the counter-examples from the \textit{context history} requires either more sophisticated Machine Learning solutions or a systematic formative study to distill new design guidelines.

Our system supports the authoring of multiple CAPs targeting different smart functions while potential errors can also be eliminated via the validation (e.g., if a CAP controls the music player state, and a user is authoring another CAP for the TV, the \textit{test case} generation algorithm would suggest the user considering the music player if these two object state \textit{context factor}s are highly correlated. 
Yet, when the user creates multiple CAPs to control the same smart function (e.g., play different genres of music under different contexts), how to inform the user of the potential conflicts would be an issue, which has also been addressed by prior works \cite{trimananda2020understanding,miandashti2020empirical}.
One straightforward solution is to directly include the other CAPs that control the same function into the \textit{test case}s, and let the user to edit all the CAPs accordingly. 
Leveraging AI-based approaches such as associate rule mining \cite{agrawal1993mining} and decision trees to enable users to manage the priority among existing CAPs could be another solution.

Most of the users agreed that the current trigger-action metaphor and the provided types of the \textit{context factor}s fulfilled their needs of authoring CAPs. 
Some users mentioned considering time-sensitive contexts such as `I was having a meeting, then make a cup of coffee'. Since a user's \textit{context history} is collected sequentially, it is feasible to calculate the uncertainty coefficients between any two \textit{context factor}s in three temporal domains: past, present, and future.
In this way, the system could enable users to author and validate time-sensitive CAPs.
Meanwhile, by using the immersive authoring environment, prior arts \cite{wang2020capturar,cao2019ghostar} visualize a user's past activities with XR animations. Instead of showing static XR icons, we could further show dynamic \textit{test case}s indicating time elapses.

\textbf{Different levels of immersiveness.}
While being situated in the XR authoring environment, the users welcomed the capability to test each case by acting it out immersively.
One user mentioned that: \textit{``Currently, everything is inside one home environment, I was considering some contexts like `go-to-office' or `while-driving'.'' (P11)} 
In addition, P2 raised an interesting feature that \textit{``I was wondering if I could see more than one test case at the same time. Maybe showing me some miniature layouts.''}
Considering the diversity of the available contexts and the different user backgrounds, providing different levels of immersiveness to visit \textit{test case}s and author CAPs would be necessary.
For instance, prior works have shown the capability to immerse users into different virtual scenes \cite{qian2022scalar,xia2018spacetime}, or even showing different room layouts at one time \cite{oberhauser2022vr} using conventional non-XR UIs. While the current system could be adapted to desktop/mobile devices (e.g., visualize CAPs with 2D UIs, visualize multiple test cases with duplicated room layouts, and use an on-screen controller to move the virtual camera within the environment during testing), further studies are required to investigate whether the benefits of such adaptation would compensate for the reduced effectiveness of the \emph{feedforward} actions for understanding the \textit{test case}s. Especially, for novice users who are not familiar with the ideas of under/over-specified CAPs and unit tests, we need more studies to evaluate whether the conventional-UI-based design would introduce additional mental loads.

\textbf{Lower the friction of using the system.}
We received complimentary feedback on the immersiveness enabled by the in-situ placed XR contents and the intuitive system operations.
Some users (P1 and P3) mentioned visualization issues when some XR contents were overlaid and clustered from some specific view directions. Meanwhile, we envision more smart objects and functions would be available in future smart home environments. 
To reduce users' visual loads, adaptively displaying the 3D contents when users move closer to them \cite{lages2019walking} or pay attention to different contents \cite{lindlbauer2019context,pfeuffer2021artention}, and adding a filter function to solely show contexts of a specific type of object would be a future improvement of the system.
Furthermore, \textit{test case}s generated by the system received positive feedback during the user study. Enabling users to act out each \textit{test case} facilitates them to understand whether the CAP needs to be modified. With these highlighted features being kept, we could add pre-processed automation to further reduce users' workloads.
Ranking \textit{test case}s according to the uncertainty coefficients would be one improvement according to P9's suggestion. Users could pay more attention to those scenarios that would more likely happen in real life, and rapidly walk through the \textit{test cases} that are listed at the end. We could also design a more advanced criterion that measures the importance of the generated \textit{test case}s (e.g., if a user edits the CAP based on a \textit{test case}, to what extent could the current CAP perform more accurately).




\section{Conclusion}

We presented \oursystem{}, a novel workflow that supports an end-user to validate the CAPs with diverse \textit{test case}s via \emph{feedforward} simulation in XR.
We first identified that existing authoring processes can result in \textit{under-specified} or \textit{over-specified} CAPs that cause unexpected behavior, leading to annoyance and frustration.
In order to address this issue, we proposed an `author-test-refine' workflow by leveraging the unit test approach in the software programming area.
In specific, using the pervasively collected everyday context record and adopting the frameworks proposed by prior context-aware application validation works, we designed a computational approach to generate multiple unit \textit{test case}s that not only are tailored to the user's personal routines but also help reveal mistakes in the authored CAPs. 
Then, while being immersed in an XR-based authoring environment, users can experience \textit{test case}s through in-situ visualization that conveys \emph{feedforward} contextual scenarios, enabling intuitive validation of CAPs.
A user study was conducted to evaluate the effectiveness of our \textit{test case} generation algorithm and the design and functionality of the XR authoring environment. 
The high accuracy of user-authored CAPs and positive feedback on system features validated the overall performance and usability of the system.
As an increasing amount of complex contexts can be digitalized into the digital space, we believe that our work can inform and provide inspiration for future investigation on how authoring systems can assist non-expert users to create error-free intelligent automation and policies that enhance the quality of life and work.


\bibliographystyle{ACM-Reference-Format}
\bibliography{references}




\end{document}